\documentclass{acm_proc_article-sp}

\begin{document}

\title{Modeling and Analysis of Scholar Mobility\\ on Scientific Landscape
\titlenote{To appear in BigScholar, WWW 2015}}
%
%
%
%
%

\numberofauthors{3} 
%
\author{
%
%
\alignauthor
Qiu Fang Ying\\
       \affaddr{Department of Information Engineering}\\
       \affaddr{The Chinese University of Hong Kong}\\
       \email{qfying@ie.cuhk.edu.hk}
\alignauthor
Srinivasan Venkatramanan\titlenote{The work was done when Dr. Srinivasan was affiliated with IE, CUHK.}\\ 
       \affaddr{Virginia Bioinformatics Institute}\\
       \affaddr{Virginia Tech}\\
       \email{vsriniv@vbi.vt.edu}
\alignauthor
Dah Ming Chiu\\
       \affaddr{Department of Information Engineering}\\
       \affaddr{The Chinese University of Hong Kong}\\
       \email{dmchiu@ie.cuhk.edu.hk}
}

\maketitle
\begin{abstract}
Scientific literature till date can be thought of as a partially revealed landscape, where scholars continue to unveil hidden knowledge by exploring novel research topics. How do scholars explore the scientific landscape , i.e., choose research topics to work on? We propose an agent-based model of topic mobility behavior where scholars migrate across research topics on the space of science following different strategies, seeking different utilities. We use this model to study whether strategies widely used in current scientific community can provide a balance between individual scientific success and the efficiency and diversity of the whole academic society. Through extensive simulations, we provide insights into the roles of different strategies, such as choosing topics according to research potential or the popularity. Our model provides a conceptual framework and a computational approach to analyze scholars' behavior and its impact on scientific production. We also discuss how such an agent-based modeling approach can be integrated with big real-world scholarly data.
\end{abstract}

\category{H.1}{Information Systems}{Models and Principles}
\category{J.4}{Computer Applications}{Social and Behavior Sciences}


\keywords{mobility; research topics; agent-based model.} 

\section{Introduction}
An important aspect of scholarly life is to do research and generate knowledge. The scholarly world can be viewed as a community effort in exploring the space of knowledge. A well-established approach for analyzing research activities is to study the networks arising out of the research process; more specifically, the (author-author) collaboration networks~\cite{newman2001structure}, and the (paper-paper) citation networks~\cite{yu1965networks}. These studies let us see how researchers collaborate, and how new knowledge discovery is built on past discoveries, and even how collaborations help knowledge discovery. But these research output based studies do not focus on motives for how researchers choose what to work on. In this work, we hypothesize that how researchers choose topics is the key hidden variable that drives the research process of the community as a whole.  An author chooses his collaborators based on similarity of topics, and a paper cites other papers on related topics.  Through researchers' strategic behavior in choosing topics to work on, the authors cause topics to emerge, grow, sustain or decay. Thus, by understanding scholars' behavior of choosing topics, we can better characterize the evolution of scientific research.

We take inspiration from two existing streams in the literature. First, Map of Science~\cite{borner2010atlas,boyack2014creation} is a well-known effort towards characterizing the relationships between different topics, using textual and citation information in the literature. By classifying and visualizing the existing papers as objects in an abstract 2-D space, they provide an atlas of science~\cite{borner2010atlas}.  If research topics and knowledge can be represented as a terrain, then researchers' choice of research topics over time can be considered as navigating through this space. The other inspiration comes from the spatial-temporal models in evolutionary biology~\cite{fletcher2006evolutionary}, where species occupy a fitness landscape that evolves subject to environmental changes. Could knowledge exploration by scholars be understood in terms of similar kind of processes, with strategic choices and motives? In the research context, the fitness landscape ideally represents the potential knowledge to be gained from different topics, which in turn drives the occupancy distribution~\cite{borner2009visual}. Scholars may choose their future topics in their current neighborhood through random search (mutation), based on their potential (selection) or by observing other scholars' behavior (imitation), and through their activities alter the fitness landscape~\cite{scharnhorst2001constructing}.


The underlying framework is essentially game-theory, where scholars are assumed to be rational decision makers following different strategies driven by certain professional or personal reward system.  However, in order to derive analytically tractable solutions for the equilibrium reached by these games, it is necessary to make strong simplifying assumptions on the models. An alternative approach is to build a computational agent-based model (ABM), where agents act according to a local set of rules, and the system's evolution and equilibrium can be simulated. ABMs have been proposed for studying various phenomena in the scientific domain such as division of labor, co-evolution of citation and collaboration networks~\cite{weisberg2009epistemic}. Furthermore, ABMs can naturally include a spatial component, thus enabling us to verify the model via a visual representation.

There are several existing ABMs that can be adapted to describe scholar mobility. For instance, in the Sugarscape model (originally proposed to study wealth distribution)~\cite{epstein1996growing}, agents move randomly around the grid, collecting ``sugar'' and ``spice''(resources that agents need to survive). Similarly in the scientific context, scholars work on different topics and gain scientific achievements, which support them to survive in academic community. However, this model only considers the simple case of random walk, and the setting of renewable resource is not suitable for our purposes. Another model by Weisberg and Muldoon~\cite{weisberg2009epistemic} describes scholars' activity as a hill climbing process to find peaks, which investigate the efficiency of three widely used searching strategies of scholars: \emph{controls}, set a direction leading to a larger height; \emph{followers}, only choose from patches visited by others; and \emph{mavericks}, only choose unvisited patches. Since there is no resource collecting in this model, agents don't have states of ``birth'', ``death'', and ``survival''. A more realistic model should incorporate both concept of resource collecting and consumption, as well as a mixture of strategies. Thus, we propose the Research Topic Selection model (RTS). We also discuss how our study can be validated against observed scholar mobility of topics derived from real-world data.

\section{The {\subsecit RTS} model}
A typical ABM consists of dynamically interacting rule-based agents, the systems within which agents interact can represent complex real-world-like scenarios. We first give definitions of the Research Topic Selection model(RTS) model in terms of elements: space, agents and rules, and then present simulation results.

\subsection{Scientific Landscape}
In our model, we consider the whole corpus of knowledge to as a map. A specific research topic is represented as an \emph{m}-dimension point on the map, in which the spatial distance between any two topics indicates their semantic correlation. 
 Although such a map could potentially be embedded in a high-dimensional space, we use a 2-D space in our study, in order to simplify the simulation and facilitate visualization.

In such a map, each point $(x_i,y_i)$ has another dimension ``height'' $h(x_i,y_i )$, used to represent the scientific significance of the corresponding topic. We call this kind of scientific map the \emph{scientific landscape}~\cite{borner2009visual}. In our model, we assume that each topic has an intrinsic scientific value before being explored, which will be revealed by scholars' research activity. In the real scientific community, the scientific value may be inferred via surrogate metrics, e.g., citation counts, research grants, etc. The problem of how to precisely measure the significance needs further study, but here we assume a scientific landscape with virtual significance.

\subsection{Scholar Agents}
In the ABM, scholars are represented as individual agents. An initial position is assigned to each agent on the scientific landscape. At each cycle of time during the simulation, agents choose a topic in the neighbourhood following certain local strategies. Upon arriving at a new research topic, agents collect a certain amount of scientific significance on it. In the scientific context, ``movement'' corresponds to scholars changing research topics; ``collected significance'' corresponds to scientific production, such as publishing papers or making breakthrough discoveries. Since a scholar leaves the community if he fails to produce research achievements for a long time, we assume there is a ``metabolism'' rate which causes each agent's collected significance to decay over time. Therefore, agents need to collect significance sufficiently to sustain in the community. In order to make the model more realistic, we set several variables for each agent:

\emph{Vision}: A scholar doesn't know the whole landscape, but can only ``see'' within a limited scope. Vision limits agents' scope of cognition hence neighbourhood to choose from. We believe that most scholars change topics over time with some continuity, since it is risky to step into a domain far away from his expertise and background. But the vision size can have a large variance among different scholars.

\emph{Academic Age}: It's not the physiological age of scholars, but the academic career length. In the real-life academic community, a scholar sooner or later retires and stops publishing papers. We can thus set ``retirement age'' for scholars. Though we set a constant retirement age for this study, it can be easily extended to a random one in the future work.

\emph{Metabolism Rate}: the decay rate of agents' collected significance in each time cycle. The terminology is adopted from Sugarscape~\cite{epstein1996growing}. One can also find supporting evidence from bibliometrics, where it is a common phenomenon for a paper's citation count to reach its peak soon after publication, and then decline steadily~\cite{wang2013quantifying}. In our model, we assume an author's significance declines exponentially, where $\lambda$ is the metabolism rate:
$$S(t)=S_0.e^{(-\lambda t)}, ~~\Delta S(t)=-\lambda.S(t)$$

\emph{Knowledge Discovery Rate (KDR)}: we assume the free significance $\Delta h_i$  an agent can collect from topic $i$ at time \emph{t} depends on its remaining significance, where $\alpha$ is the KDR:
$$\Delta h_i=\alpha .h_i (t)$$

One advantage of the RTS model is that it is flexible to add or modify variables. However, it is to be noted that a small change in a variable may lead to significantly different behaviors.

\subsection{Strategies}
The Weisberg and Muldoon model~\cite{weisberg2009epistemic} categories searching strategies by how agents consider topics' intrinsic research value and popularity to make decisions. We agree that intrinsic research value and popularity are two major factors influencing scholars' choices. But scholars in our model have varing goals, e.g., to survive long in the academic community or achieve greater scientific significance, not just finding the most important topics. Thus we define four types of strategy by different criteria:\\
\textbf{Experts}: \emph{research value} - prefer topics with high scientific value. They are scholars who have capability to recognize topics' intrinsic value and make decisions independently.\\
\textbf{Followers}: \emph{hotness} - prefer currently hot topics. They follow trends because they lack ability to estimate the value of topics as experts, or are interested in immediate rewards.\\
\textbf{Mavericks}: \emph{novelty} - prefer unexplored topics. These authors are often pioneers discovering the ``new world'', though the strategy is potentially risky.\\
\textbf{Conservatives}: \emph{maturity} - prefer well established research areas (not necessarily trending).

Agents employ the following movement rules at each cycle:\\
\textbf{Experts}:\\
\emph{Check}: Any patch in my vision has higher significance than my current patch?\\
\underline{If yes}, move to the patch of highest significance; \underline{If no}, stay.\\
\textbf{Mavericks}:\\
\emph{Check}: Any patch in my vision hasn't been visited yet?\\
\underline{If yes}, randomly move to one of the candidate patches.\\
\underline{If no}, \emph{check}: any patch has higher significance?\\
\indent\underline{If yes}, move to one of those patches; \underline{If no}, stay.\\
\textbf{Followers}:\\
\emph{Check}: Any patch in my vision has other visitors currently?\\
\underline{If yes}, check: any of them has higher significance?\\
\indent\underline{If yes}, randomly move to one of those patches;\\
\indent\underline{If no}, \emph{check}: any other patch is currently unoccupied?\\
\indent\indent\underline{If yes}, randomly move to one of patches; \underline{If no}, stay.\\
\underline{If no}, randomly move to one of patches.\\
\textbf{Conservatives}:\\
\emph{Check}: Any patch in my vision has ever been visited?\\
\underline{If yes}, \emph{check}: any of them has higher significance?\\
\indent\underline{If yes}, randomly move to one of the patches;\\
\indent\underline{If no}, \emph{check}: any other patch hasn't been visited?\\
\indent\indent\underline{If yes}, randomly move to one of patches; \underline{If no}, stay.\\
\underline{If no}, randomly move to one the patches.

We don't claim that they are only strategies for scholars and our model is flexible to include other strategies.

\section{Simulations}

\begin{figure}[htb]
\centering
\epsfig{file=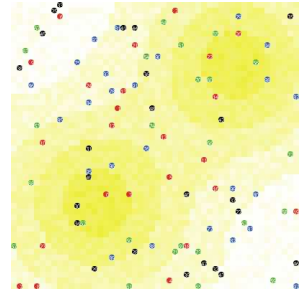, width=1.5in,} 
\caption{\label{fig.1}2D representation of scientific landscape with four types of agents randomly located.}
\end{figure}

We conduct simulations of RTS model using the tool NetLogo. In our simulations, the scientific landscape is built on a 50 patch by 50 patch grid. we use two Gaussian functions to define the significance~\cite{weisberg2009epistemic} with added noise, as in Fig.~\ref{fig.1}. In the beginning, each scholar agent $a_k$ is assigned to a patch $(x_{a_k},y_{a_k})$ ,  with initial collected significance (or wealth) $r_o (a_k)$, academic age  $g_o(a_k)$, and vision $v(a_k)$. Then in each cycle time, agents move to one patch and collect significance on the patch. Before we discuss the simulation results, we first define several metrics for the performance evaluation of strategies.
\subsection{Evaluation Matrices}
In our model, science significance is a non-increasing resource which will be consumed by scholar agents. On the other hand, agents will depart the community if they deplete their wealth or reach the retire time. Thus, we are most interested in questions:
	a) What type of scholars can survive the longest?
    b) What type of scholars can achieve the most success?
	c) Which strategy is the most social efficient? We design the following three evaluation metrics:
\begin{enumerate}
    \itemsep=-5pt
	\item \textbf{Individual Cumulative Achievements (ICA)}: summation of achievements an agent collected during the whole academic life.
	\item \textbf{Progress}: ratio of consumed significance by all agents compared to the landscape's initial significance.
	\item \textbf{Coverage}: ratio of patches visited as least once by all agents.
\end{enumerate}

Agents ICA evaluates strategies by the efficiency for individual, progress and coverage evaluate the social efficiency for the entire community.

\subsection{Single Strategy Scenarios}
\begin{figure}[htb]
\centering
\epsfig{file=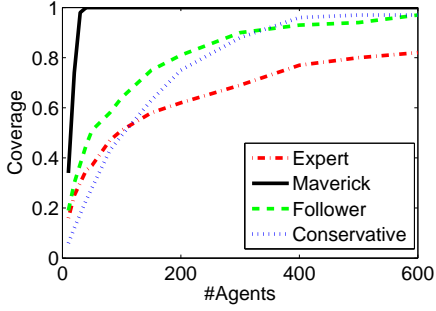, width=2.4in,} 
\caption{\label{fig.2}Coverage of single strategy at $t=100$.}
\end{figure}

\begin{figure}[htb]
\centering
\epsfig{file=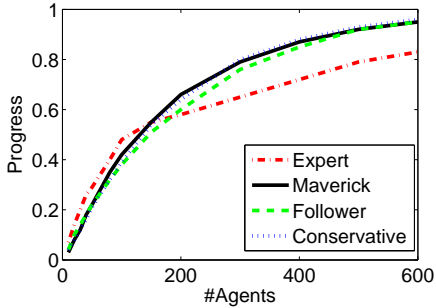, width=2.4in,} 
\caption{\label{fig.3}Progress of single strategy at $t=100$.}
\end{figure}

We first ran simulations involving single type of strategy in the population. As can be seen in Fig.~\ref{fig.2}, mavericks lead largest coverage, experts produce lest coverage. The cause is that experts tend to congregate on high topics, while other three types of agents are flexible to choose diverse topics, especially mavericks. From Fig.~\ref{fig.3}, we find experts has largest progress when the population contains less than 150 agents . But when the population increases, since all experts still choose topics of highest significant, they have intense competition, resulting in low Progress.

\begin{figure}
\centering
\epsfig{file=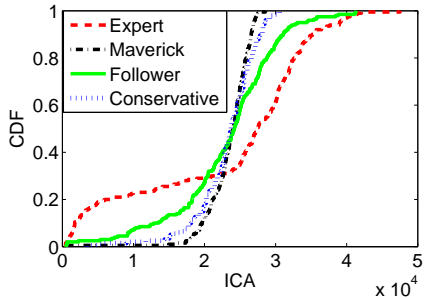, width=2.4in,} 
\caption{\label{fig.4}ICA distribution of single strategy at $t=100$.}
\end{figure}

Fig.~\ref{fig.4} show the personal ICA distribution, we can see that experts have skewed ICA, while mavericks have balanced ICA. It gives an interesting implication: in a community of mavericks, agents tend to study diverse topics and have balanced personal achievements; however, in a community of experts, since each agent are able to identify the highest significant topics, there are fierce competition among them which leads to skewed individual ICA.\\

\subsection{Multiple Strategies Co-existing Scenarios}
We then conduct simulations with a mixed population of four types of agents. An interesting question is which type will win? Will the four types of agents collaborate or compete with each other? A population of 200 agents are initialized in the system, equally assigned to the four types. We study the influence of three parameters: vision, metabolism rate and KDR on the performance of four strategies

\subsubsection{Vision}
\begin{figure}[htb]
    \begin{minipage}[t]{0.4\linewidth}
    \centering
    \epsfig{file=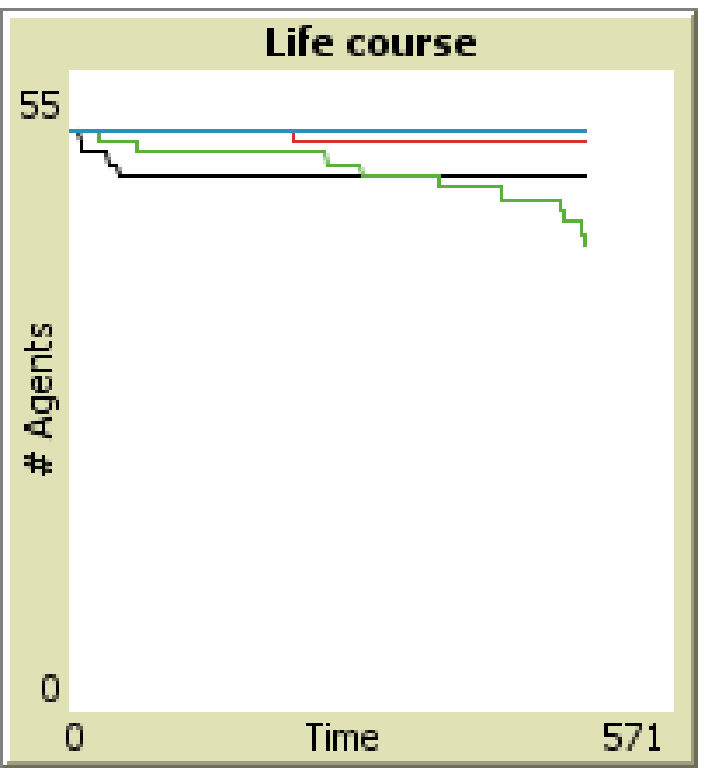, width=1.3in,} 
    \end{minipage}
    \begin{minipage}[t]{0.4\linewidth}
    \centering
    \epsfig{file=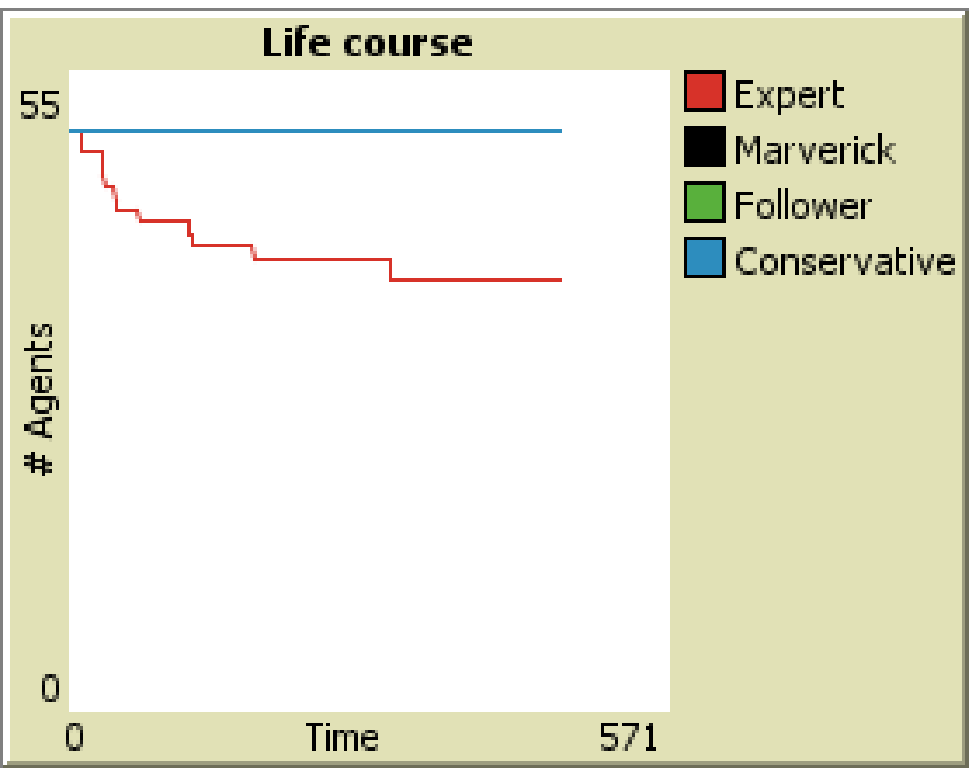, width=1.8in,} 
    \end{minipage}
\caption{\label{fig.5}life course of agents in mixed population with vision=1 and vision=10.}
\end{figure}

\begin{figure}[htb]
\centering
\epsfig{file=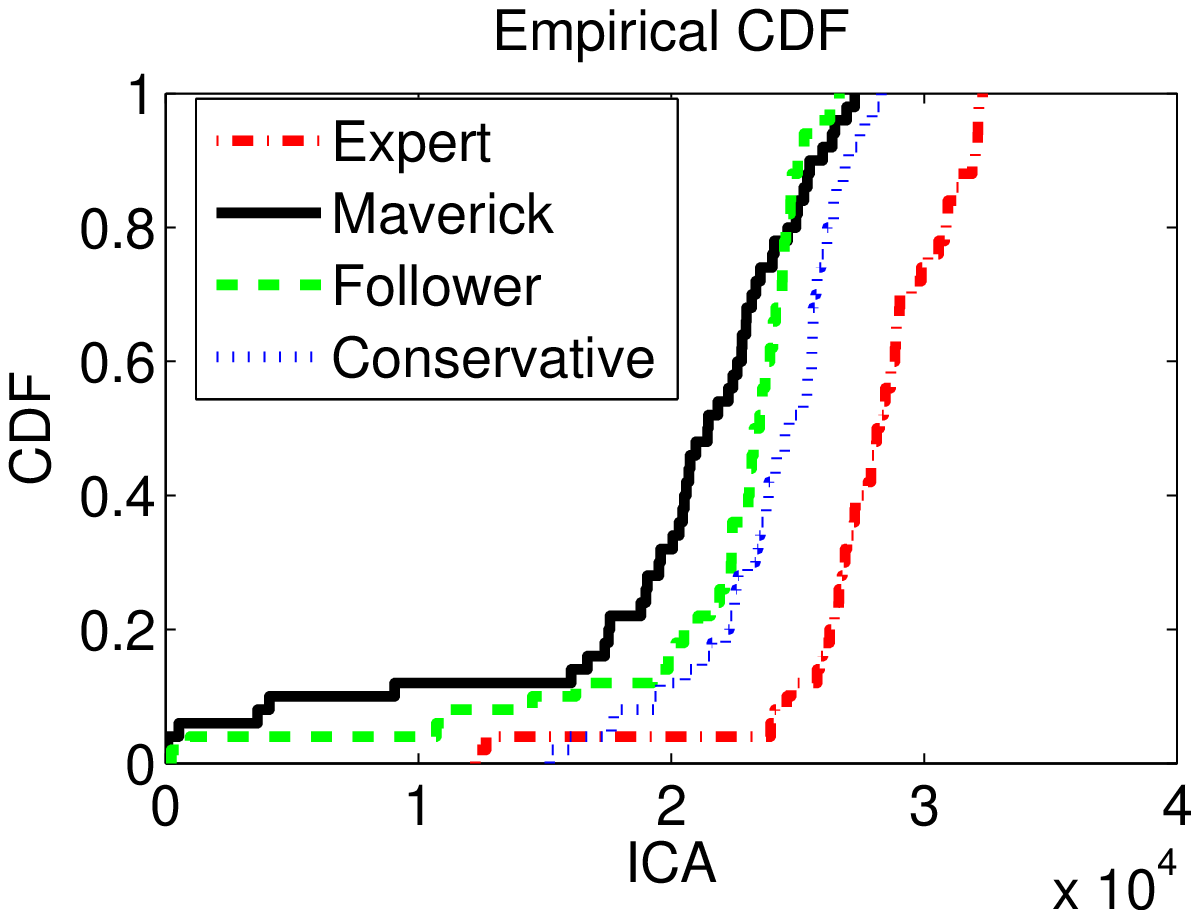, width=2.4in,} 
\caption{\label{fig.6}ICA distribution of mixed strategies with vision=1.}
\end{figure}

\begin{figure}[htb]
\centering
\epsfig{file=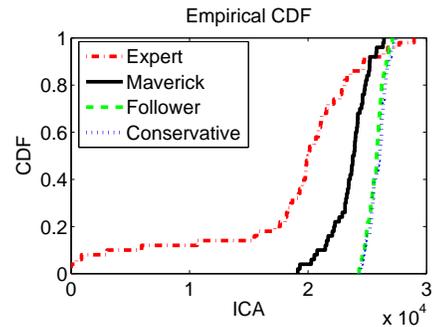, width=2.4in,} 
\caption{\label{fig.7}ICA distribution of mixed strategies with vision=10.}
\end{figure}

Vision defines the size of neighborhood an agent can see when making decisions. From Fig.~\ref{fig.5}, Fig.~\ref{fig.6} and Fig.~\ref{fig.7}, we find a counterintuitive phenomenon that with large size of vision, experts become less suited for survival than in case with small size of vision. The reason lies on the social impact of the crowd. If the source of information is large enough, followers, conservatives and mavericks can utilize the wisdom of crowd to make good decisions.  On the contrary, the larger space experts can see, the higher possibility that all of them aggregate to the high positions, which results in intense competition. This phenomenon proves the importance of utilizing wisdom of crowd in a social community.

\subsubsection{Metabolism Rate}
\begin{figure}[htb]
    \begin{minipage}[t]{0.4\linewidth}
    \centering
    \epsfig{file=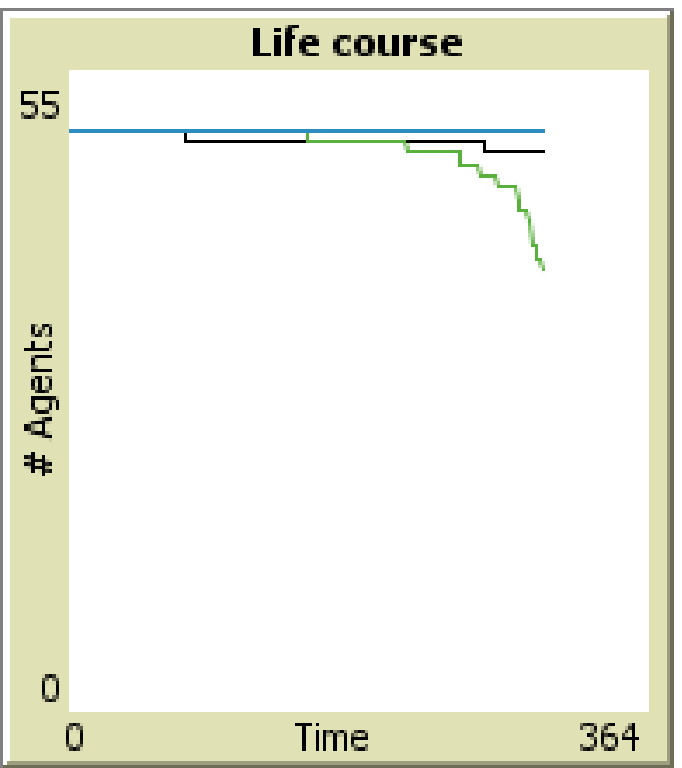, width=1.3in,} 
    \end{minipage}
    \begin{minipage}[t]{0.4\linewidth}
    \centering
    \epsfig{file=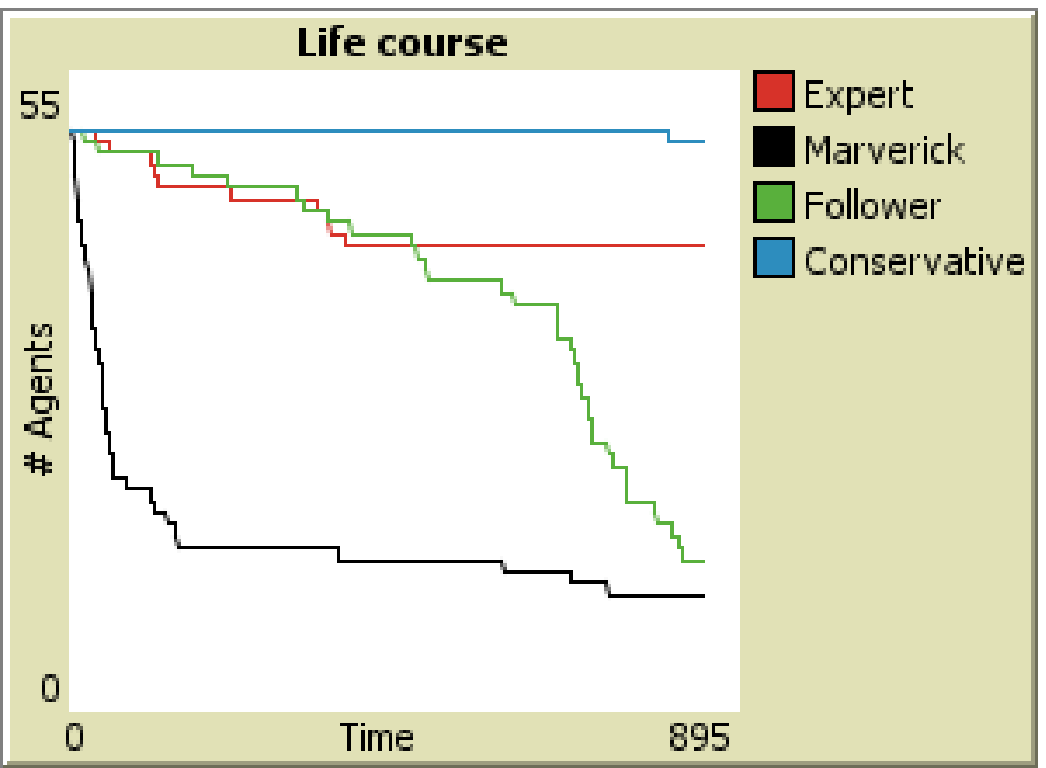, width=2.0in,} 
    \end{minipage}
\caption{\label{fig.8}life course of agents in mixed population with metabolism rate $\lambda=0.2$ and $\lambda=0.8$.}
\end{figure}

\begin{figure}[htb]
\centering
\epsfig{file=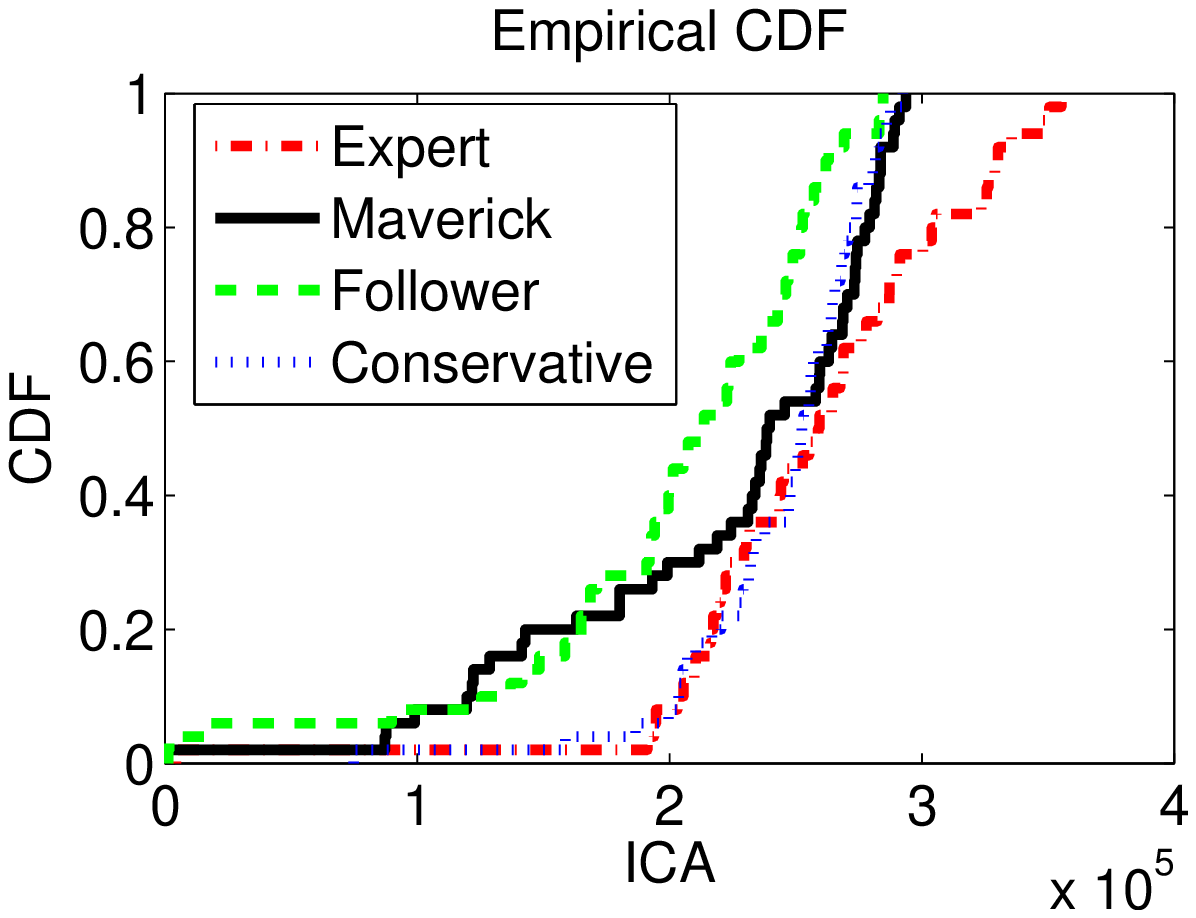, width=2.4in,} 
\caption{\label{fig.9}ICA distribution of mixed strategies with metabolism rate $\lambda=0.2$.}
\end{figure}

\begin{figure}[htb]
\centering
\epsfig{file=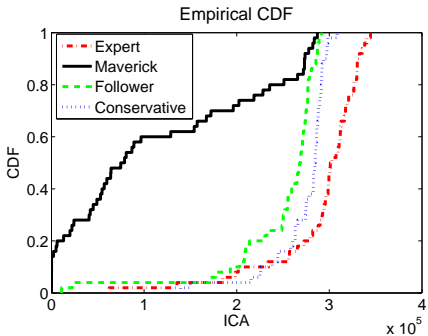, width=2.5in,} 
\caption{\label{fig.10}ICA distribution of mixed strategies with metabolism rate $\lambda=0.8$.}
\end{figure}

Metabolism Rate $\lambda$ determines how quickly agents' achievements decay over time. From Fig.~\ref{fig.8}, Fig.~\ref{fig.9} and Fig.~\ref{fig.10}, first we find that when the metabolism rate becomes larger, the probability of agents departing the system at early age also increases, leading to skewed ICA distributions. Moreover, we find that the influence of metabolism rate is most severe on mavericks. Mavericks have small number of departures in case of $\lambda=0.2$, but the early departure increase dramatically when $\lambda β=0.8$, much worse than other three. Therefore, mavericks prefer low metabolism rate, which in reality corresponds to a relaxed research environment, e.g, adequate faculty positions, sufficient research resource and funding, no demanding requirement of yearly publications. Historically, one can see that several breakthrough discoveries were made in such less demanding environments, which foster innovation.

\subsubsection{Knowledge Discovery Rate}
\begin{figure}[htb]
    \begin{minipage}[t]{0.4\linewidth}
    \centering
    \epsfig{file=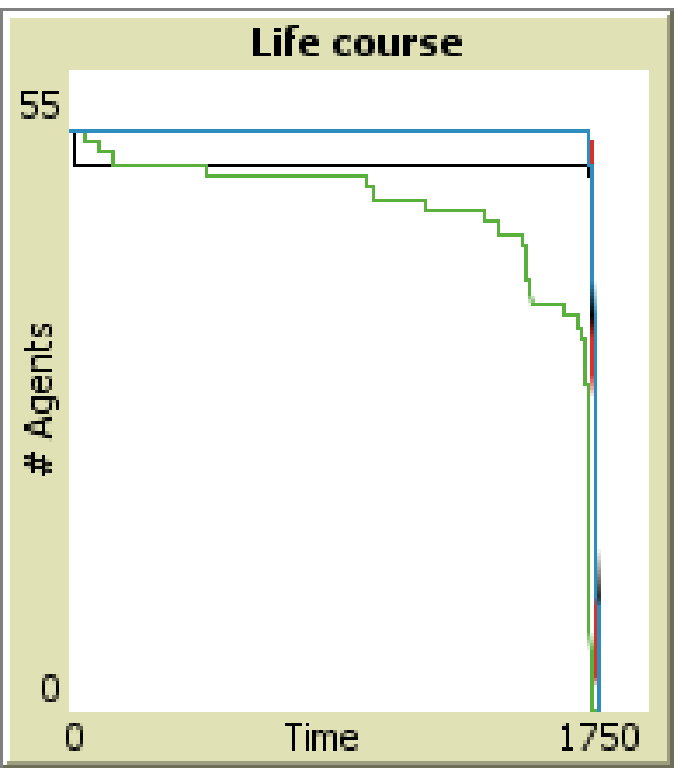, width=1.3in,} 
    \end{minipage}
    \begin{minipage}[t]{0.4\linewidth}
    \centering
    \epsfig{file=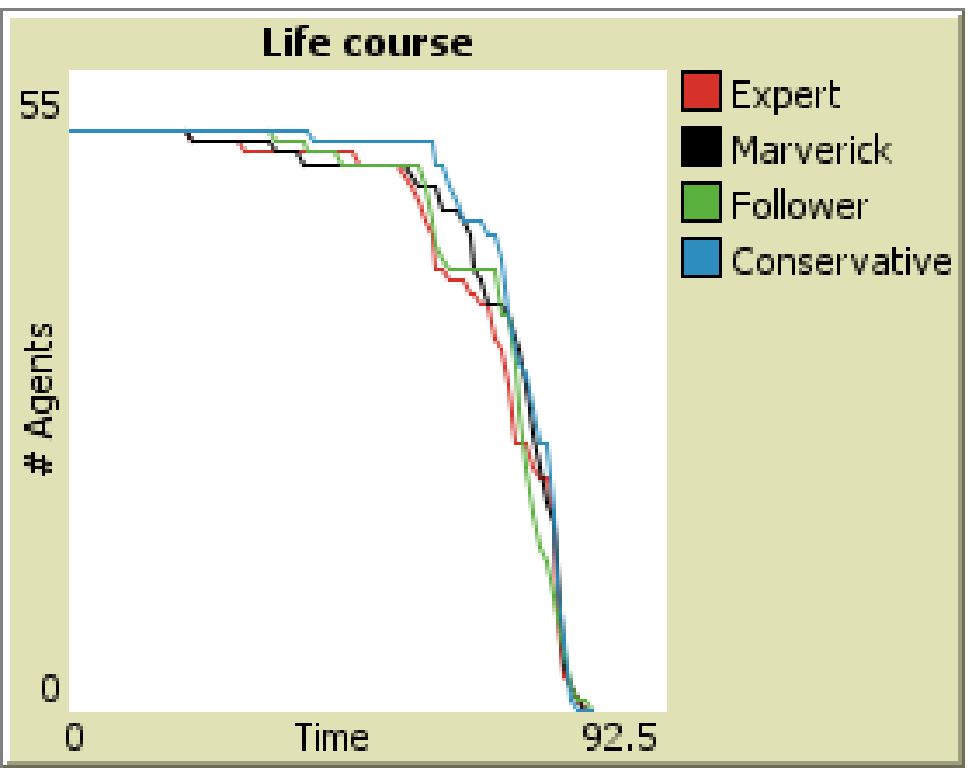, width=1.85in,} 
    \end{minipage}
\caption{\label{fig.11}life course of agents in mixed population with KDR $\alpha=0.05$ and $\alpha=0.5$.}
\end{figure}

\begin{figure}[htb]
\centering
\epsfig{file=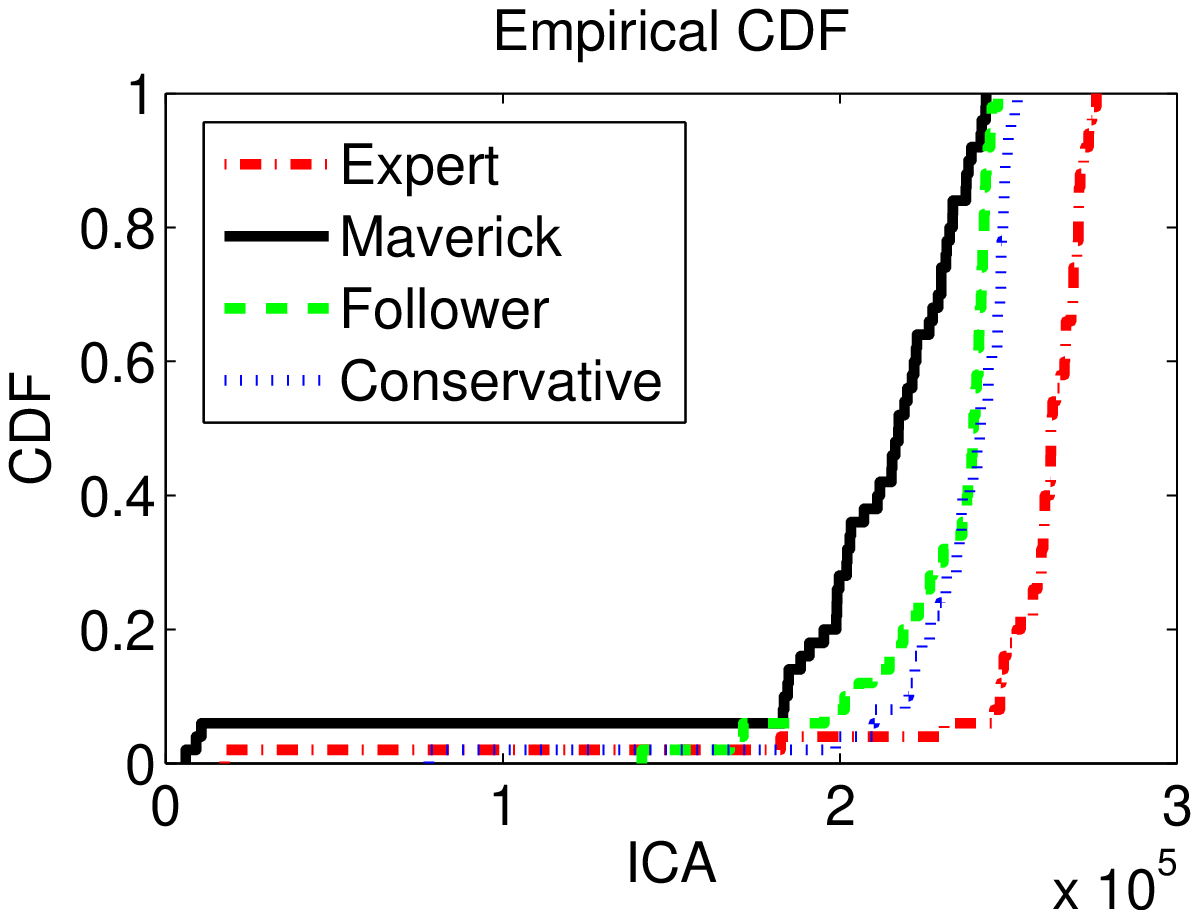, width=2.4in,} 
\caption{\label{fig.12}ICA distribution of mixed strategies with KDR $\alpha=0.05$.}
\end{figure}

\begin{figure}[htb]
\centering
\epsfig{file=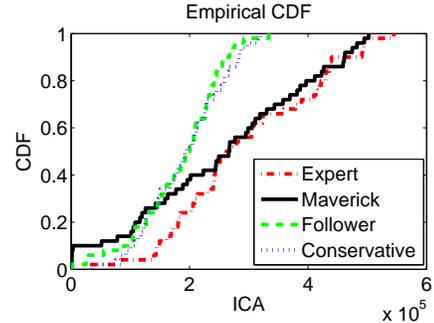, width=2.4in,} 
\caption{\label{fig.13}ICA distribution of mixed strategies with KDR $\alpha=0.5$.}
\end{figure}

As we can see from Fig.~\ref{fig.11}, Fig.~\ref{fig.12} and Fig.~\ref{fig.13}, mavericks are most sensitive to the change of Knowledge Discovery Rate $\lambda$. A large $\lambda$ means that the significance of a topic will be quickly consumed, thus the early arrivals to an unexplored area can have a big advantage. Since mavericks always seek novel and less visited topics, they are more likely to take the advantage of being pioneers.

From above analysis, we can have an insight of what roles the four types of agents play in the scientific community. From the perspective of personal success, the four types of agents have different sensitivity of the parameters in research circumstance. From the view of social efficiency, both mavericks and experts make essential contributions: experts act as leaders to rich areas around their expertise, and mavericks explore innovative areas to maximize the system's diversity. Followers and conservatives sustain and build on the research efforts of mavericks and experts, collectively increasing the community's knowledge base. This inspire us to do further investigation of the structure and component of the community, by analysis on big scholarly data.

\section{Discussion \& Future work}
\textbf{\emph{Model Extension:}}
In our RTS model, we study a scenario that each scholar agent adopts a fixed type of strategy during the simulation. But in real academic community, a researcher might change strategies in different stages of her career, e.g., being a follower as a fresh graduate, and switching to be an expert after accumulating enough knowledge and expertise. In other words, a scholar chooses the best strategy, whose utility changes with his academic age, vision, and also the strategies by other scholars. The current RTS model can be considered as an abstraction of how things play out given a particular mix of strategies in a certain equilibrium. Extending our model to allow changing strategies will be an interesting future work. Furthermore, one can also introduce arrivals and departures for agents in this dynamic process. Another key component missing in this work, is that of collaborative efforts. Since the model requires extensive modifications to accommodate collaboration, we do not discuss it in this paper.

\textbf{\emph{Topic Mapping and Visualization:}}
We assume a 2-D scientific landscape of multiple Gaussian functions in the simulation, but what is the real scientific landscape? Building the map of scientific landscape from real scholarly data is a big challenge.

Significant efforts build the map of science by using the textual and citation information in the literature. Topics model~\cite{fortuna2005visualization}, e.g., LDA , dimensionality-reduction techniques~\cite{borner2010atlas,borner2009visual,boyack2014creation}, e.g., PCA and LSA are used to derive a low-dimensional representation of publications. More recently, there is renewed interest in using non-linear dimensionality reduction (NLDR) techniques such as Deep Learning  ~\cite{ranzato2008semi} to improve the accuracy of visual representation.

\begin{figure}[htb]
\centering
\epsfig{file=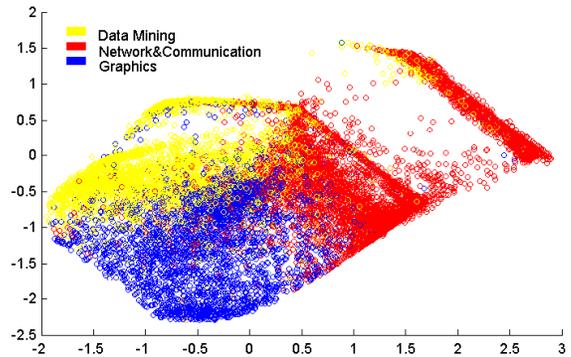, width=3.8in,} 
\caption{\label{fig.14}visualization of papers using 2D code by a 3857-100-10-5-2 autoencoder.}
\end{figure}

To have a qualitative evaluation whether the neural network method works, we conduct an simple experiment by implementing both PCA and autoencoder algorithms~\cite{hinton2006reducing} on a sample data from Microsoft academic Libra dataset.The data consists of 15606 papers from five conferences in the computer science domain, as listed in Table ~\ref{tab:papercount}. Using titles and keywords of the papers, we built a corpus of 3857 words. To facilitate visualization, the output data is reduced to 2 dimensions. Fig.~\ref{fig.14} shows the resulting 2D map of papers using the autoencoder algorithm. What is a good map of science? Minimally, the structure of the map must show strong clustering of papers from known scientific subdomains. From Fig.~\ref{fig.14}, we find the map produced by autoencoder does show a clear division of the three subdomains we picked, and even distinguishes the different conferences inside the same subdomain. By comparison, our attempt with the PCA algorithm had a more ambiguous partition (not shown due to lack of space). Thus, the neural network approach seems promising. In the future work, we plan to study how to train neural networks using complete Libra dataset to build the scientific landscape.

\textbf{\emph{Validation model with big scholarly data:}}
While most spatial ABMs~\cite{payette2012agent} are used for modeling only, we hope to achieve some level of validation of the RTS model with real big scholarly data, by visualizing the research topic trajectories of different scholars, and analyzing the same authors using traditional methods to assess their performance, in terms of activity, productivity, and life time.
\begin{table}[htb]
\centering
\caption{Information of 5 conferences in CS}\label{tab:papercount}
\begin{tabular}{|c|c|c|}
\hline
Venue & Domain & \#papers \\
\hline
KDD  & Data Ming(DM)  & 2038 \\
\hline
ICDM & Data Ming(DM)  & 2166 \\
\hline
INFOCOM & Network \& Communication(NC) & 6027 \\
\hline
SIGCOMM & Network \& Communication(NC) & 1223 \\
\hline
SIGGRAPH & Graphics(GR) & 4152 \\
\hline
\end{tabular}
\end{table}

\section{Conclusions}
We proposed a computation model to study the behavior of individual scholars on how they choose research topics in their research career. From preliminary simulation results, we find even simple models can reflect interesting phenomena of research practices in scholarly communities.  We created four types of scholars who play different roles: experts lead scholars to topics with high research potentials, mavericks are the pioneers of novel topics, followers and conservatives utilize the wisdom of crowds. The ratio of scholars adopting certain strategies has significant impact on the health and progress of our scientific community. Validating our model with big scholarly data is a challenging direction in our future study.







%
\bibliographystyle{abbrv}
\bibliography{BigScholar-qf}  
%
%

\balancecolumns
\end{document}